\documentstyle[12pt,titlepage,fleqn]{article}

     \newcommand{\be}{\begin{equation}}
     \newcommand{\ee}{\end{equation}}
     \newcommand{\ba}{\begin{eqnarray}}
     \newcommand{\ea}{\end{eqnarray}}

     \newcommand{\oh}{\frac{1}{2}}
     \newcommand{\om}{\omega}
     
     \newcommand{\p}{\hat{p}}
     \newcommand{\cO}{{\cal O}}

\begin{document}

\begin{titlepage}
\hfill{ Alberta Thy-25-95}
\vspace*{0.1truecm}
\begin{center}
\vskip 1.0cm
{\large\bf  QUANTUM SOLITONS LEAD TO YUKAWA COUPLING}

\vskip 2.0cm

{\large F. ALDABE\footnote{E-mail: faldabe@phys.ualberta.ca}}

{\large\em Theoretical Physics Institute\\
University of Alberta\\
Edmonton, Alberta\\
 Canada, T6G 2J1}
\vspace{.5in}

\today \\

\vspace{.5in}
{\bf ABSTRACT}\\
\begin{quotation}
\noindent
Recently, it was shown that zero modes in semiclassical soliton models
do not lead to Yukawa couplings.
We show that taking into account the contributions of the quantum
soliton into the renormalization scheme, which cannot be done in
semiclassical treatments, leads to a Yukawa coupling.
A similar analysis should be possible for the Skyrmion, renewing the
hope, that this model will lead to a correct description of hadron physics.

\end{quotation}

\end{center}
\vfill
\end{titlepage}
One of the mayor drawbacks of semiclassical treatments of
soliton models in describing
nuclear physics at intermediate energies has been
the lack of a Yukawa couplings between the fluctuations and the solitons. The
minimization of the potential in order to define the soliton yields a theory
without an explicit coupling.  Many attempts have been made \cite{ohta} to
show that the Yukawa coupling is implicit in the spurious sector generated
by the symmetry breaking solution.
However, recently \cite{yo4},  with the formalisms
presented in \cite{todo}
it has been shown that the
treatments of the zero modes in \cite{ohta} are incorrect because
the asymptotic states used to derive the Born terms that imply the existence
of a Yukawa coupling are unphysical; when physical asymptotic states
are used there are no Born terms to leading order which will insure the
existence of such a coupling.

Here we make use of a quantum description of the soliton and ``renormalize" the
theory taking into account the contributions of the soliton which is now
treated as a particle.  This renormalization procedure cannot be carried
out in a semiclassical treatement as the baryon is not a quantum particle.
We will demonstrate that the integration of the energetic mesons
in the quantum soliton model leads to
the existence of a Yukawa coupling obtaining a well defined theory
which can describe the basic features of baryon physics.
Our attention has been drawn away
from the connection between the integation of high frequency
mesons and the Yukawa coupling because
the semiclassical renormalization scheme cannot be
applied to the Skyrme model.
However, our method still apply to this model.
The result we will obtain  here are hinted in \cite{man}.  There,
the mesons were integrated out completely yielding an effective soliton
Hamiltonian: a massive Thirring model where the solitons interact strongly.
Here we will carry a partial integration of the meson field.
As the partial integration involves only energetic
mesons, we may speak of a ``renormalization" scheme,
which may also be applied to the Skyrme model.

There are certain  features of the semiclassical treatment of
soliton models which we outline here.
There is a classical solution with nonvanishing topological
charge which is interpreted as the baryon number.  All
the fluctuations, but a zero mode, have
the physical interpretation of mesons about
this solution.  These fluctuations
 cannot be plane waves.  Otherwise the fluctuations are not
physical, due to the presence of the zero mode,
 in contradiction with our knowledge of mesons \cite{yo4}.
{}From this fact, it follows that there is no Yukawa coupling between the
mesons and the baryons which would imply that these models are not
adequate to describe hadron physics.  However, there is also another
inconsistency.  There is no natural cutoff scale for the mesons.
This means that these models are ill defined twice.

The sine-Gordon Hamiltonian can be written as
\be
H=\int\{\p^2+\phi^{'2}+ V\}
\ee
with $V$ admiting solitonic solutions.
The opertor which measures the  topological
charge of the solution is defined as
\be
Q_t=\int j^0=
\beta\epsilon^{0\nu}\int\partial_{\nu}\phi=
\beta(\phi(x=\infty)-\phi(x=-\infty)).
\ee
Where $j^0$ is the zeroth component of the topological current,
and $\beta$ is a parameter which is realted to the mass of the solution.
It also determines
the coupling of the soliton to the mesons as we shall demonstrate.
The classical value, or equivalently, the vacuum expectation value of $Q_t$
is related to the baryon number.
It also holds that the topological charge is conserved in the
vacuum because
\be
i{[H,Q_t]}= \beta(\p(x=\infty)-\p(x=-\infty))\label{cos}
\ee
has vanishing expectation value.  The statements in this paragraph
hold regardless of
whether the soliton is classical or quantum in nature.

Classical solutions to the sine-Gordon Hamiltonian with nontrivial
soliton number carry mass $M(\beta)>>1$.  Renormalization of the mass
has been carried in \cite{dv} for such classical treatments,
and shown to be independent of the cutoff,
and of $\cO(M^{-1})$.

I will now construct a soliton vertex operator associated to
the above Hamiltonian.
We impose that the soliton operator $\sigma(x)$ which is a function of
the fields satisfying the sine-Gordon equation of motion
\be
\sigma(x)=:e^{-i \beta^{-1}\int^x_{-\infty}{\p}}:\label{sigma}
\ee
be an eigenvector of the soliton currents.
It is easy to check that a soliton
state created by $\sigma(x)$ is normalized because
\be<vac|\sigma^+(x)\sigma(x)|vac>=<vac|vac>=1
\ee
Since we associate the
topological current to the soliton current we have
\be
{[j^{0}(y),\sigma(x)]}=\sigma(x)\delta(x-y)\label{com}.
\ee
The topological charge of the state created by $\sigma(x)$
will be
\be
<vac|\sigma^+(x)Q_t\sigma(x)|vac>=1\ee
The topological charge of the soliton operator is one.  The topological charge
is also conserved in time since
\be
<vac|\sigma^+(x)
\beta(\p(x=\infty)-\p(x=-\infty))\sigma(x)|vac>=0.\ee
Using the fact that
\be
<vac|(H-H)\sigma(x)|vac>=0
\ee
and demanding that
\be
H\sigma(x)|vac>=M\sigma(x)|vac>+\cO(1)
\ee
it follows that
\be <vac| \sigma(x)|vac>=0.\label{dec}
\ee
Therefore, the state created by the soliton vertex operator
is orthogonal to the vacuum to leading order.

As expected in classical soliton models
after the creation of a
soliton, the field $\phi$ must increase at the left boundary
by a certain amount \cite{man}
$\beta^{-1}$, the right boundary being unchanged.
This is indeed the case since
\be
{[\phi(x) ,\sigma(y)]}=\beta^{-1}\sigma(y)\theta(y-x).\label{profi}
\ee
Therefore in the soliton vacuum created by $\sigma(x)$
the field does not have  vanishing expectation value.
 Moreover, it is of ${\cal O}({\beta^{-1}})$ as
in semiclassical models.  It does follow that the field so defined
has the same properties of a semiclassical soliton solution.
It breaks translational
invariance,  its energy is finite and the square root of
its energy density is a localized
distribution,
the delta function,
rather then the distribution of  a point particle.  Moreover, it may not be
constructed perturbatively.

Having defined the soliton state created by the soliton vertex operator
we now study the spectrum of
its excitations, the mesons,
 and its
ground state energy with respect to the bare vacuum.  To do
this we make use of a transformation of the normal modes which
will annihilate the 1 soliton vacuum, their conjugates will
create a phonon when acting over the one soliton vacuum.  These
procedure is similar to the construction of the quasiparticles in the
BCS model.
The operators which annihilate the soliton vacuum are
\be
\gamma_n=-c_n+a_n\;\;\;\;c_n=-i\beta^{-1}\sqrt{\frac{\om_n}{2}}\int^x_{\infty}
\psi_n^*\in {\bf C}.
\ee
where $a_n$ is an anihilation operator in which the fluctuations of the
classical theory can be expanded.
The constants $c_n$ are non vanishing only because the integral is over
a fraction of space.

Thus the Hamiltonian on this vacuum reads
\be
H=\sum_n(
\om_n|c_n|^2+\om_n(c_n\gamma_n^++h.c.)+\om_n(\gamma^+_n\gamma_n+\oh))+\cO(\beta).\label{hf}
\ee
We could have obtained this expression for the Hamiltonian if we would
have considered the transformation
\be
\sigma^+(x) H\sigma(x)
\ee

The first term in (\ref{hf}) is the energy of the one soliton vacuum.  It
is not finite since we have no cutoff for the meson spectrum.  If this
hamiltonian is going to make sence as one which describes a baryon and
the mesons about the baryon, the first term better be finite.  On the other
hand,
as mentioned before, the renormalization of the classical Hamiltonian
leads to corrections to the mass of $\cO(M^{-1})$.  In order for
the description of the classically renormalized soliton model to
be compatible with that of the quantum description, both must yield
the same one soliton vacuum energy.  This is achieved only when we
introduce a cutoff $\Lambda$ such that
\be
M=\sum_n^{\Lambda} \om_n|c_n|^2,
\ee
Thus, effectively renormalizing the quantum soliton model.

Then, the Hamiltonian (\ref{hf}) should be
\be
H=\sum_n^{\Lambda}(
\om_n|c_n|^2+\om_n(c_n\gamma_n^++h.c.)+\om_n(\gamma^+_n\gamma_n+\oh))+\cO(\beta).
\label{hfr}
\ee
This means that the quantum description of the soliton we have used
effectively integrates out mesons whose energy is greater than $\Lambda$.
This integration of energetic mesons is done through a different procedure
than that
used
in the semiclassical treatment because in the quantum case, the baryon
is treated as a particle which actively takes part in the renormalization
scheme (through its construction as a coherent states of mesons)
while it is not treated as such
in the classical description.  This implies that the renormalization schemes
can lead to different physical picture.

The presence of a cutoff was already
expected from the classical field profile (\ref{profi}): in order to
transform the classical solution to that of (\ref{profi}) we need to
integrate out mesons whose wave number are of the order of the extension
of the classical solution of the semiclassical soliton model which is
or the order of the meson mass. As expected.

The second term in
(\ref{hfr}) cannot be obtained in the classical treatment of solitons, even
after renormalization.
This term, is the so much looked for Yukawa coupling which yields the
matrix element of the appropiate order
\be
<\sigma(x)|H|\gamma^+_k,\sigma(x)>\sim \sqrt{M}
\ee

The last
term is just the single particle Hamiltonian for the mesons.
Notice that the spectrum of the meson modes remains unaltered.
There are also
no higher point interactions leading to divergences
signalining that the energetic
mesons have
effectively been integrated out to all orders of perturbation theory.

{\bf Acknowledgments}

I would like to thank the ICTP for kind hospitality during the early stages
of this work.

\pagebreak

\end{document}